\documentclass[aps,prc,a4paper,preprintnumbers,superscriptaddress,showpacs,amsmath,floatfix,groupedaddress,nofootinbib,twocolumn]{revtex4}
\usepackage{hyperref}
\usepackage{multirow}
\usepackage{color}
\usepackage{tabularx}
\usepackage{graphicx}
\usepackage{amsthm}
\usepackage{amsmath}
\usepackage{amssymb}
\usepackage{bm}
\usepackage{textcomp}
\usepackage{url}
\usepackage{tabularx}
\usepackage{verbatim}

\graphicspath{ 
  {Plots/} 
}
\newcommand{\pt}{\mbox{$p_T$}}

\newcommand{\auau}{\mbox{Au+Au}}

\newcommand{\pp}{\mbox{{\it p+p}}}

\newcommand{\snn}{\mbox{$\sqrt{s_{_{NN}}}$}}

\newcommand{\gev}{\mbox{$\mathrm{GeV}$}}
\newcommand{\bgev}{\mbox{$\mathbf{GeV}$}}
\newcommand{\gevc}{\mbox{$\mathrm{GeV/}c$}}
\newcommand{\gevcc}{\mbox{$\mathrm{GeV/}c^2$}}

\newcommand{\rda}{\mbox{$R_{d\textrm{Au}}$}}
\newcommand{\rdac}{\mbox{$R_{d\textrm{Au}}^{c \rightarrow e}$}}
\newcommand{\rdab}{\mbox{$R_{d\textrm{Au}}^{b \rightarrow e}$}}

\newcommand{\mncoll}{\mbox{$\langle N_{\textrm{coll}} \rangle$}}

\newcommand{\sqrts}{\mbox{$\sqrt{s}$}}
\newcommand{\sqrtsNN}{\mbox{$\sqrt{s_{_{\mathrm{NN}}}}$}}

\newcommand{\pA}{{\it p+A}}
\newcommand{\dA}{{\it d}+\textrm{Au}}
\newcommand{\pps}{pp}

\DeclareMathOperator{\arctanh}{arctanh}

\newcommand{\ctoe}{\mbox{$c \rightarrow e$}}
\newcommand{\btoe}{\mbox{$b \rightarrow e$}}

\newcommand{\sigpp}{\sigma_{\pps}}
\newcommand{\sigppinel}{\sigma_{\textrm{inel}}}

\newcommand{\eh}{\mbox{$e^{\rm HF}$}}

\begin{document}
\title{Estimate of cold nuclear matter effects on bottom production in \dA\ collisions at $\snn=200 \ \bgev$}

\affiliation{Warsaw University of Technology, Warsaw, Poland}
\author{Daniel~Kiko\l{}a}\affiliation{Warsaw University of Technology, Warsaw, Poland}
\author{Andrzej~Lipiec}\affiliation{Warsaw University of Technology, Warsaw, Poland}

%\preprint{Intended for Phys. Rev. C, version 1}
\preprint{version 1.0}
\date{\today}
\pagenumbering{arabic}
\pagestyle{myheadings} %headings
\thanks{}
\markboth{{\small D. Kiko\l{}la, A. Lipiec}}{{\small production in $\auau$ $\snn = 200 \ \gev$}} 
%\footnote{Footnote Here}

%\linenumbers

%%%%%%%%%%%%%%%%%%%%%%   Abstract   %%%%%%%%%%%%%%%%%%%%%%%%%

\begin{abstract}

We investigate modification of the bottom quark production due to cold nuclear matter effects (CNM) at mid-rapidity in \dA\ collisions at $\snn = 200 \ \gev$ at RHIC. Our results indicate that bottom production is not suppressed due to CNM effects in those collisions. We also found that shadowing and initial $k_T$ breadboarding for charm quarks explains at low \pt\ ($\pt < 3\ \gevc$) the enhancement of  heavy flavor decay electron yield in \dA\ collisions at $\snn = 200 \ \gev$ compared to \pp.  

\end{abstract}

\maketitle

%%%%%%%%%%%%%%%%%%%%%%  Sections   %%%%%%%%%%%%%%%%%%%%%%%%

\section{Introduction\label{introduction}}

High energy heavy ion collisions provide an opportunity to create in a laboratory a Quark Gluon Plasma, QGP, a state of matter with quark and gluon degree of freedom. Charm and bottom quarks are important probes of the properties of the QGP because they are created in the initial scatterings with large momentum transfer and are expected to interact with the QGP differently than light quarks (see Ref.~\cite{Rapp:HF:review} and references therein). For instance, studies of the heavy quark energy loss in nucleus-nucleus collisions could provide information about transport properties of the created nuclear medium. 

It is important to measure charm and bottom production separately in \aa\ collisions to have a full picture of energy loss for light and heavy quarks. This was a major motivation for recently completed upgrades at the STAR and PHENIX experiments at the Relativistic Heavy Ion Collider (RHIC) at Brookhaven National Laboratory. These upgrades include a micro-vertexing detectors: Heavy Flavor Tracker (HFT) at STAR and  Silicon Vertex Tracker (VTX) and Forward Silicon Vertex Detector (FVTX) at PHENIX, which allow measurement of charm and bottom production. Charm will be measured via direct reconstruction of hadronic decays of D mesons. Electrons from semi-leptonic decays of bottom hadrons (noted here as $\btoe$) are the most feasible tools for bottom studies. STAR and PHENIX collected large data samples of Au+Au collisions at $\snn = 200 \ \gev$ which will allow precise measurement of heavy quark production and their nuclear modification factors. For interpretation of these results, it is important to have an estimate of so-called cold nuclear matter (CNM) effects for c and b quarks i.e. modification of production not related to the QGP formation. 

Experimentally we address these effects by measuring particle production in \pA\ or \dA\ interactions. Such data for b and c quarks are not available so far (charm and bottom separation in \pA\ will be possible in 2016, after \pA\ run at RHIC). However, it is crucial to have an estimate of CNM effects on bottom quark production when the first precise Au+Au data are available in 2015.

Moreover, current data for electrons from semi-leptonic decays of heavy flavor hadrons, $e^{HF}$, show an enhancement of the production in central and minimum bias \dA\ collisions at mid-rapidity at RHIC~\cite{Phenix:NPE:dAu}. Recent observations of collective behavior of light hadrons in \dA\ collisions at RHIC and \pA\ at LHC triggered speculations that this enhancement is an indication of collective phenomena (radial flow) for heavy quarks in \dA~\cite{HF:flow:dAu}. However, this enhancement could be also owing to the CNM effects. 

In this paper we estimate the modification of the bottom quark production due to cold nuclear matter effects at top RHIC energy. First, we make a minimal set of assumptions about those effects for charm and we simulate electrons from charmed meson decays (\ctoe) in \dA\ reactions. We consider initial transverse momentum ($k_T$) broadening of partons and modification of the parton distribution function in a nucleon in a nucleus compared to a free proton (so called shadowing). Then we simulate $c \rightarrow e$ in \dA\ with those CNM included using measured charm \pt\ spectrum in \pp\ collisions at $\sqrts = 200 \ \gev$ as an input. Then we subtract \ctoe\ contribution from \eh\ yield measured by PHENIX collaboration to obtain electrons from bottom hadron decays.
We also investigate if the $e^{HF}$ enhancement can be explained by established cold nuclear matter effects namely $k_T$ broadening and shadowing. 

\section{Simulation setup \label{simulation}}

\begin{figure}[hbtp]
\begin{center}
\includegraphics[width=.45\textwidth]{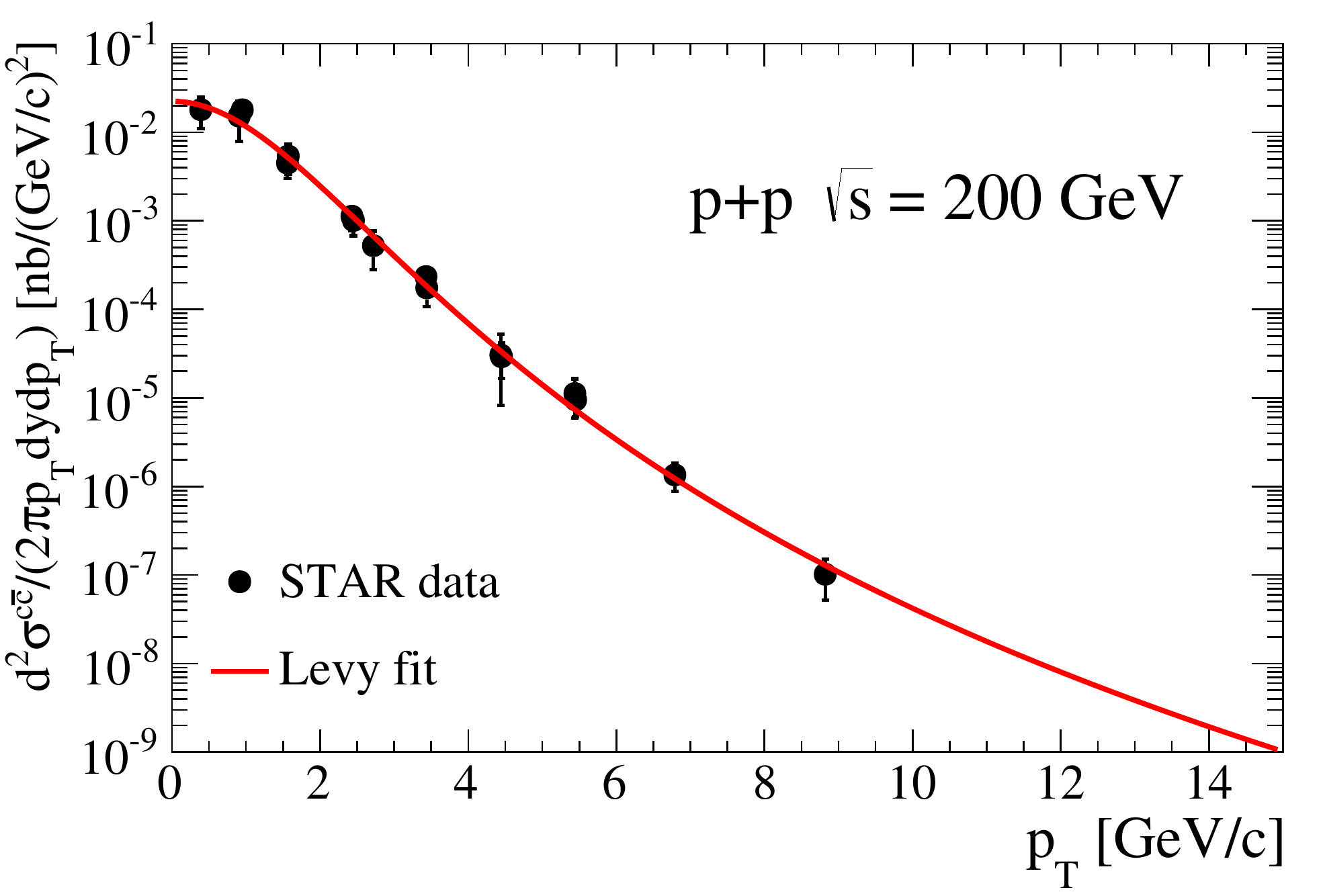}
\put(-50,80){$|y|<1$}  
\caption{
  (Color online) Differential charm quark cross-section \pp\ collisions at $\sqrts = 200 \ \gev$ (combined published STAR data~\cite{STAR:charm:pp} and preliminary results\cite{STAR:charm:QM2014}) with a Levy function fit. The error bars represent statistical and systematic uncertainties added in quadratures.
}
\label{fig:charm_pp}
\end{center}
\end{figure}

\begin{figure}[hbtp]
\begin{center}
\includegraphics[width=.45\textwidth]{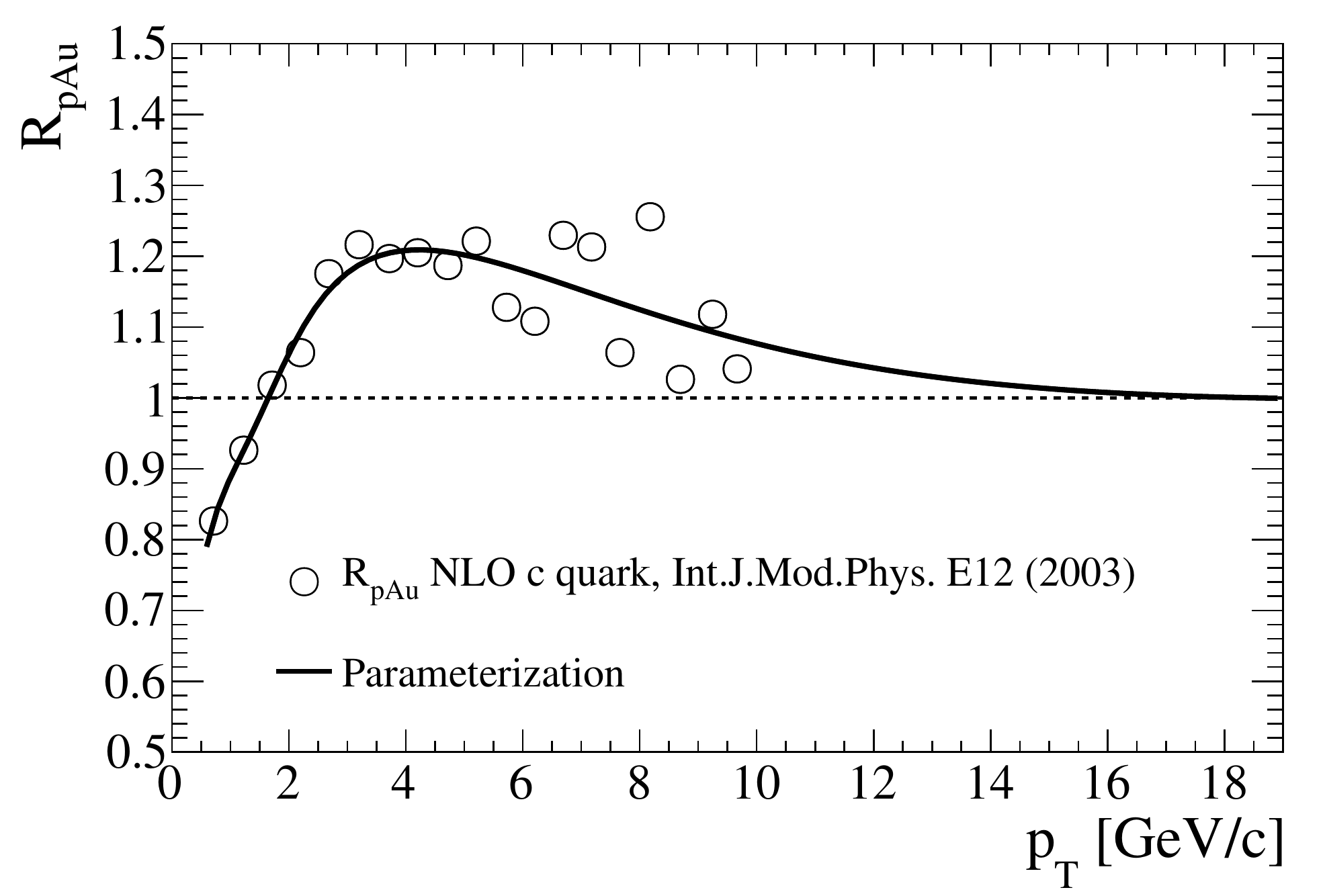}    
\caption{
  (Color online) Predictions for charm quark $R_{pA}$ as a function of \pt~\cite{RpA:theory:Vogt} (open symbols) with the parametrization $R_{pA}(\pt) = A \tanh(Bx)+C \arctanh(Dx) + E \exp(-x) +  F \exp(-x^2)$ used in this work. 
}
\label{fig:RpA}
\end{center}
\end{figure}

We use charm differential cross section in \pp\ collisions at $\sqrts = 200 \ \gev$ as input in our simulations. We construct the input spectra by combining the published STAR data~\cite{STAR:charm:pp} and recent preliminary results~\cite{STAR:charm:QM2014}. The $\pt$ spectrum is parametrized with a Levy function $f(\pt) = A \frac{(n-1)(n-2)}{nT(nT + m_D(n-2))}(1 + \frac{m_T^2-m_D}{nT})^{-n}$, where $A$, $T$ and $n$ are free parameters, $m_D = 1.86484 \ \gevcc$ is $D^0$ mass and $m_T = \sqrt{m_D^2 + \pt^2}$.
We chose this parametrization because it represents charm \pt\ spectrum in a broad \pt\ range (1-18 \gevc) in \pp\ collisions at $\sqrts = 500 \ \gev$~\cite{STAR:charm:QM2014}. The fit describes the data very well at $\sqrts = 200 \ \gev$ (Fig.~\ref{fig:charm_pp}) as well. We assume that charmed hadrons ($D^0$, $D^{\pm}$, $D*$, $D_S$) have the same shape of \pt\ spectrum and we use an average branching ratio Br = 10.5\%. We simulate a decay kinematics with PYTHIA8(version 8.8176)~\cite{Pythia8} and weigh the results according to the charm \pt\ distribution in \pp. We assume charm has a uniform rapidity distribution within $|y|<1$ and use electrons with $|\eta|<1$ to obtain the $e^{HF}$ differential cross section at mid-rapidity ($y=0$). The electron spectrum is normalized to match the charm cross-section, calculated with the Levy function, times the branching ratio. We estimate the uncertainties on the fit with a Monte Carlo method: We move points in the D-meson \pt\ spectrum within their uncertainties, assuming  that they have Gaussian distribution. Each point is shifted vertically by $N\sigma_y$, where $\sigma_y$ is an overall uncertainty for a given point (combined statistical and systematic uncertainties) and $N$ is a random number from the standard normal distribution. Then we re-fit the data to get a new \pt\ parametrization and calculate $\ctoe$ spectrum. We repeat this procedure 1000 times and obtain a distribution of the $\ctoe$ for each \pt\ bin. Standard deviations of these distributions give an estimate of uncertainty of the \ctoe\ yield due to input D-meson spectrum.

We consider two cold nuclear effects for charm quarks: broadening of initial $k_T$ distribution in \pA\ collisions (which leads to so-called Cronin effect namely an enhancement of particle production at intermediate \pt\ in \pA\ compared to \pp\ collisions) and modification of the parton distribution function in the nucleus compared to the free proton (shadowing). We use predictions from Ref.~\cite{RpA:theory:Vogt} to parametrize those effects. In those calculations EKS98 shadowing parameterization is used and the $k_T$ broadening from multiple scattering of the projectile partons in the target is parametrized as $\langle k_T^2 \rangle_A = \langle k_T^2 \rangle_p (\langle \nu \rangle -1) \Delta^2$, where $\langle k_T^2 \rangle_p$ is parton transverse momentum in \pp\ collisions, $\langle \nu \rangle$ is average number of collisions in a proton-nucleus interaction and $\Delta^2$ describes the strength of the nuclear broadening ($\Delta^2$ depends on the scale of the interactions and it is larger for $b\overline{b}$ than $c\overline{c}$ production). 

Fig.~\ref{fig:RpA} shows the original calculations~\cite{RpA:theory:Vogt} for next-to-leading order (NLO) inclusive charm quark production together with parametrization of $R_{pA}(\pt) = A \tanh(Bx)+C \arctanh(Dx) + E \exp(-x) +  F \exp(-x^2)$, where $A,B,C,D,E$ and $F$ are parameters. We obtained the parameters from fit to the predictions in Fig.~\ref{fig:RpA}. We assume those CNM effects are small at high \pt, thus we added a constrain $R_{pA} = 1$ at $\pt=20 \ \gevc$.
To obtain charm spectrum in minimum-bias \dA\ collisions, we multiply \pp\ data by $R_{pA}(\pt)$ and scale with average number of binary collisions $\mncoll = 7.5 \pm 0.4$~\cite{STAR:dAu:2003}.

\section{Results \label{results}}

\subsection{Comparison with measured \eh\ and \ctoe\ in \pp\ collisions at $\sqrts = 200 \ \gev$ }

\begin{figure*}[hbt]
\begin{center}
\includegraphics[width=.95\textwidth]{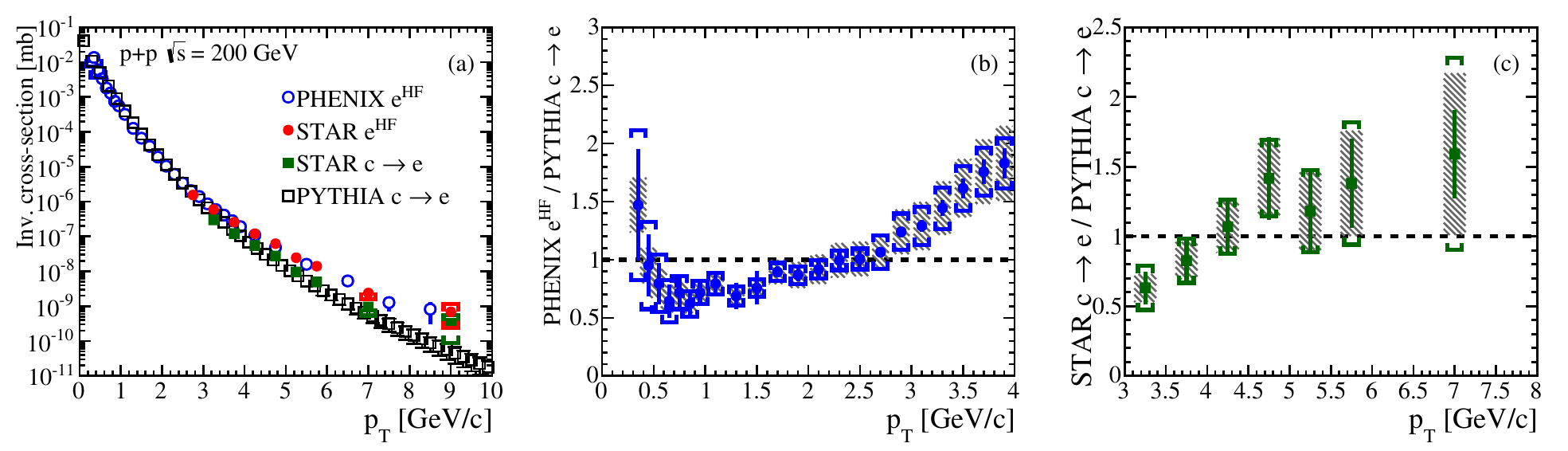}  
\caption{
  (Color online) (a) \pt\ spectrum of electrons from semi-leptonic heavy meson decays measured by STAR~\cite{STAR:NPE:pp200GeV} and PHENIX~\cite{Phenix:NPE:pp:AuAu} compared to our $\ctoe$ calculations. (b) Ratio of PHENIX \eh\ to $\ctoe$ (this work). (c) Ratio of STAR $\ctoe$ data to our $\ctoe$ results. Hashed boxes in (b) and (c) show uncertainties on our $\ctoe$ calculations.  
}
\label{fig:npe:pp}
\end{center}
\end{figure*}

We first check if our simulations reproduce experimental data in \pp\ collisions at $\sqrts = 200 \ \gev$ at mid rapidity. Figure~\ref{fig:npe:pp}(a) shows a \pt\ spectrum electrons from heavy flavor hadron decays, \eh, reported by PHENIX~\cite{Phenix:NPE:pp:AuAu} and STAR~\cite{STAR:NPE:pp200GeV}. These results include contribution both from charm (\ctoe) and bottom (\btoe) quarks. We also plot spectra for charm and bottom separately by STAR~\cite{STAR:NPE:pp200GeV} and our \ctoe\ simulations. Charm dominates \eh\ spectrum for $\pt<2 \ \gevc$ (bottom contribution is $\sim 20\%$ at $\pt = 2 \ \gevc$ and decrease with decreasing \pt~\cite{STAR:cb:separation}) thus we expect a good agreement between our results and \eh\ data at $\pt< 2 \ \gevc$. Figure~\ref{fig:npe:pp}(b) shows a ratio of \eh\ electrons and our simulations and these data agree within statistical and systematic uncertainties. However, there are deviations for $0.7 < \pt < 1.5 \ \gevc$. This difference could be due to different values of the charm cross-section reported by PHENIX and STAR. PHENIX measured  ${\rm d}\sigma^{c\overline{c}}/{\rm d}y$ via single electron spectra and obtained ${\rm d}\sigma^{c\overline{c}}/{\rm d}y = 119 \pm 12 ({\rm stat.}) \pm 38 ({\rm syst.}) \ \mu b$~\cite{Phenix:NPE:pp:AuAu} while STAR measurement using direct reconstruction gives ${\rm d}\sigma^{c\overline{c}}/{\rm d}y = 161 \pm 20 ({\rm stat.}) \pm 34 ({\rm syst.}) \ \mu b$~\cite{STAR:charm:QM2014}. Thus we expect $\sim15-20\%$ difference at low \pt\ between PHENIX \eh\ measurement and our \ctoe\ simulations using charm \pt\ spectrum from STAR. The difference can also arise due to assumptions in this work. Figure~\ref{fig:npe:pp}(c) shows a ratio of STAR \ctoe\ measurement to our results. We focus on $3 < \pt\ < 8 \ \gevc$ where data have a reasonable precision. We observe a hint of different slopes for \pt\ spectrum in the data and simulations, but overall these results agree within uncertainties.  

\subsection{Cold nuclear matter effects}

\begin{figure*}[hbtp]
\begin{center}
\includegraphics[width=.7\textwidth]{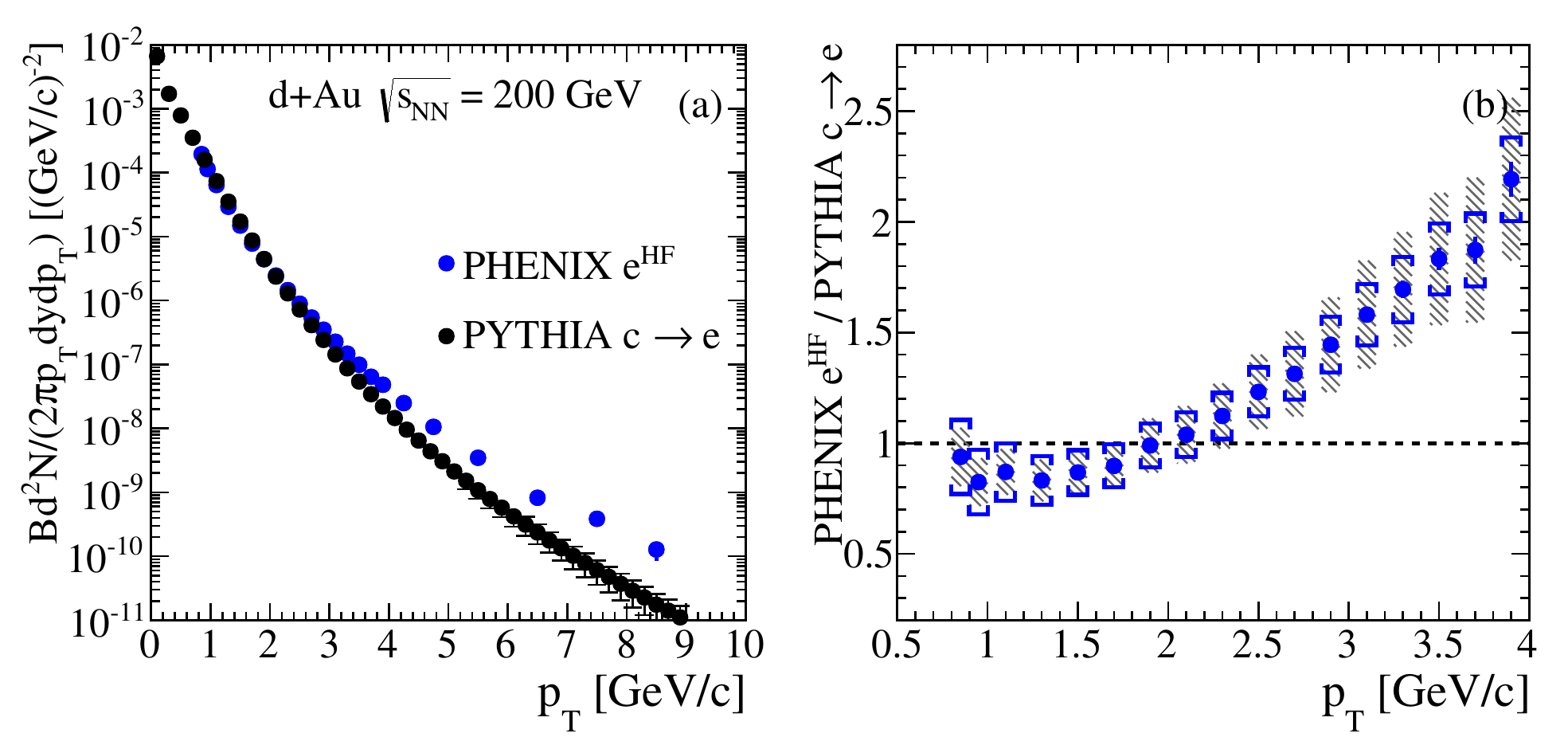}  
\caption{
  (Color online) PHENIX \eh\ spectrum in minimum-bias \dA\ collisions~\cite{Phenix:NPE:dAu} at mid-rapidity compared to our \ctoe\ calculations (a) and their ratio (b). Hashed boxes in (b) show uncertainties on $\ctoe$ calculations.  
}
\label{fig:npe:dAu}
\end{center}
\end{figure*}

\begin{figure}[!htbp]
\begin{center}
\includegraphics[width=.45\textwidth]{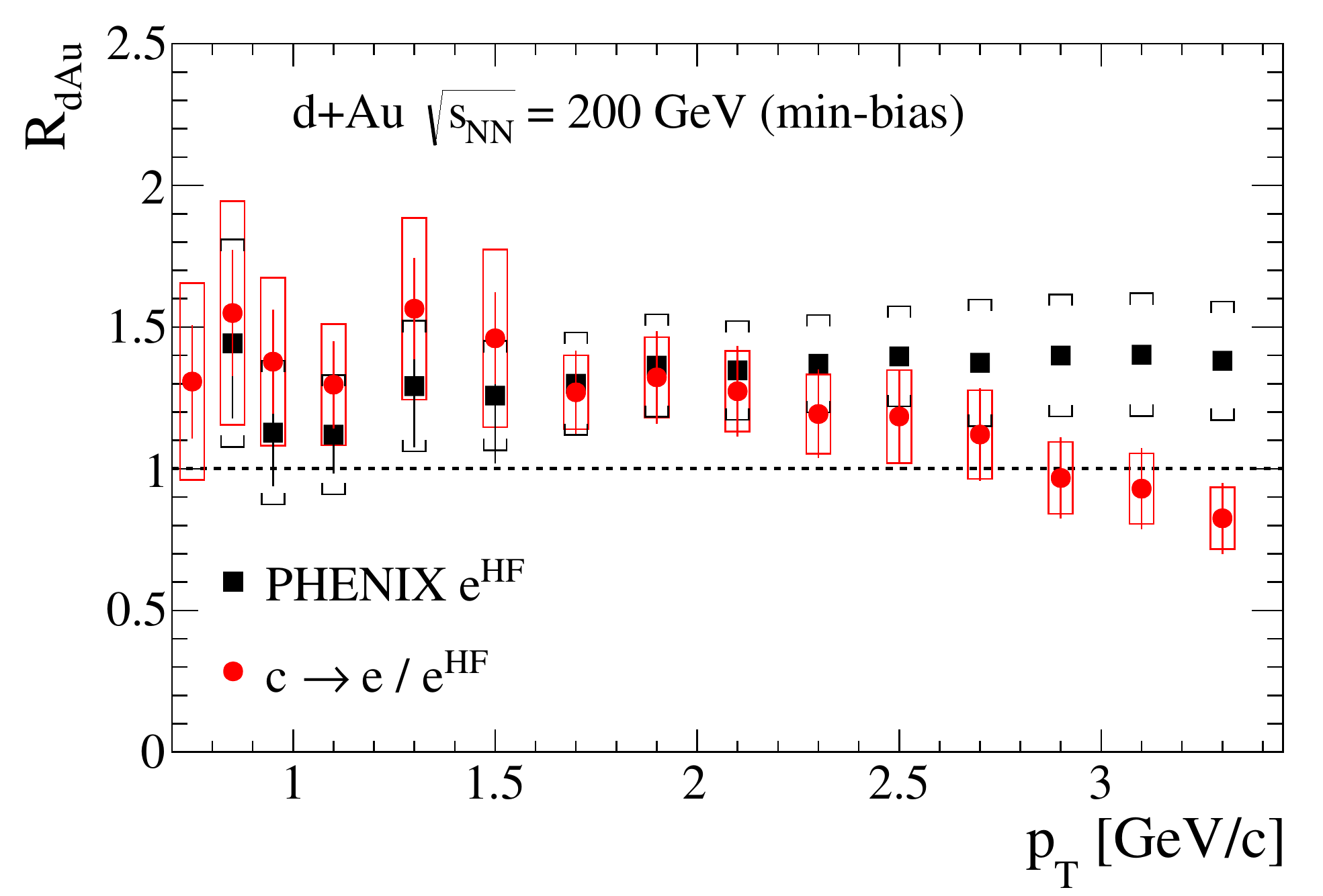}
\put(-60,40){$|\eta|<0.35$}  
\caption{
  (Color online) Nuclear modification factor for electrons from heavy flavor hadron decays from PHENIX (black squares) and ratio of \ctoe\ in \dA\ to \eh\ in \pp\ scaled by \mncoll. 
  }
\label{fig:npe:RdAu}
\end{center}
\end{figure}

Figure~\ref{fig:npe:dAu} shows \eh\ \pt\ spectrum in minimum-bias \dA\ collisions at $\snn = 200 \ \gev$ compared to our \ctoe\ calculations. Our results match the \eh\ data well for $\pt<2.5 \ \gevc$, which is expected since charm dominates \eh\ spectrum for $\pt<2 \ \gevc$. We subtracted the \ctoe\ yield from \eh\ \pt\ spectrum to obtain the \btoe\ yield. We quantify a change of \ctoe\ and \btoe\ production in \dA\ with nuclear modification factor $\rda$. $\rda$ is the ratio of the electron yield (\ctoe\ or \btoe) in $\dA$ and $\pp$ collisions, where the latter is scaled by the average number of binary collisions $\mncoll$ in $\dA$: 
\begin{equation}
\rda = \frac{\sigppinel}{\mncoll}\frac{\mathrm{d}^{2}N_{d\rm{Au}}/\mathrm{d}y\mathrm{d}\pt}{\mathrm{d}^{2}\sigpp/\mathrm{d}y\mathrm{d}\pt}
\end{equation}
where $\sigppinel$ is the inelastic cross section in $\pp$ collisions, $\sigppinel = 42 \pm 3$~mb, $N_{d\rm{Au}}$ is \ctoe\ (or \btoe) yield in $\dA$ collisions and $\mathrm{d}^{2}\sigpp/\mathrm{d}y\mathrm{d}\pt$ is the  \ctoe\ (or \btoe) cross section in $\pp$ collisions, respectively. STAR results~\cite{STAR:NPE:pp200GeV} for \ctoe\ and \btoe\ serve as a baseline and $\mncoll = 7.5 \pm 0.4$~\cite{STAR:dAu:2003}.

\begin{figure}[!htbp]
\begin{center}
\includegraphics[width=.45\textwidth]{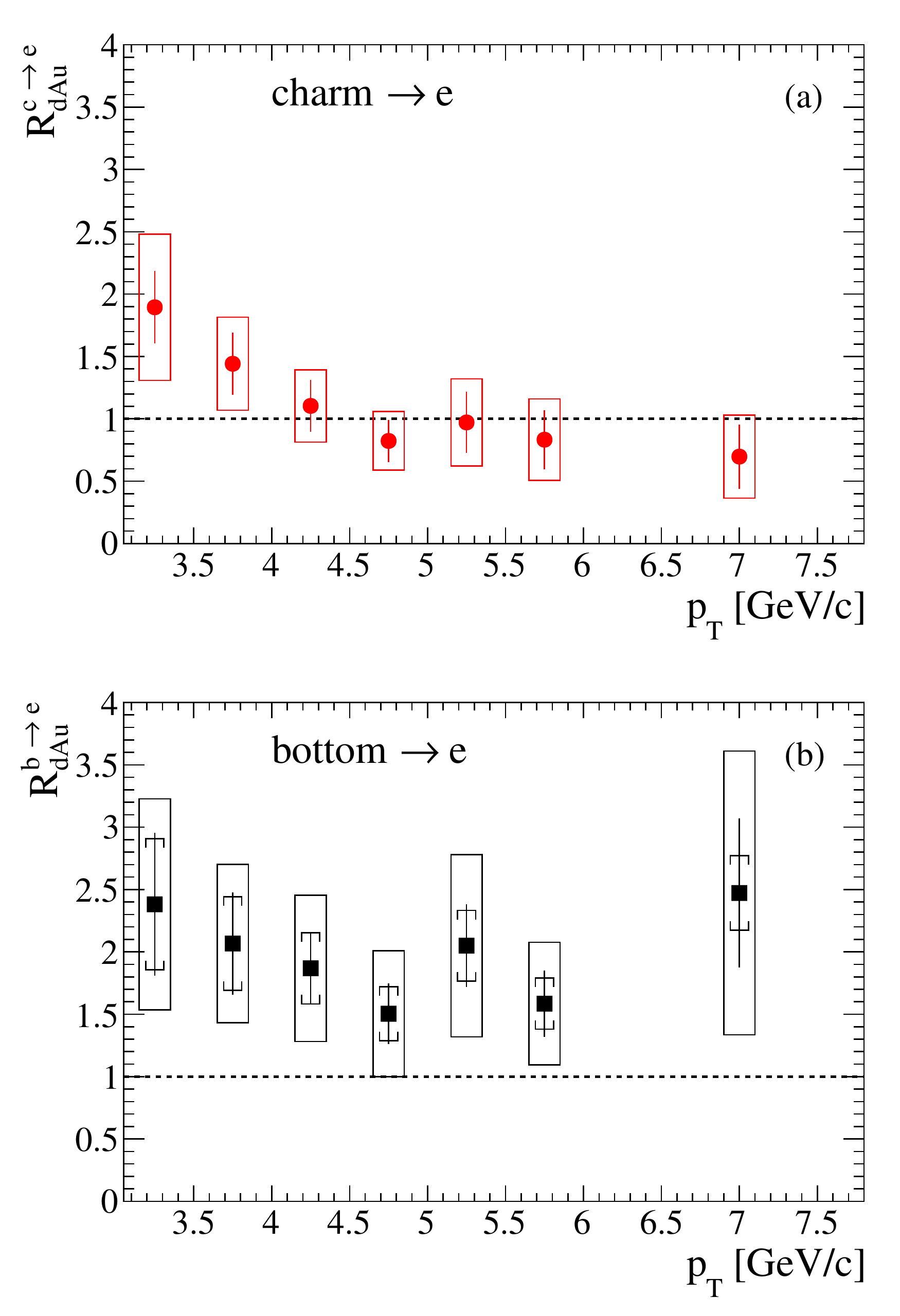}  
\caption{
  (Color online)  Nuclear modification factor for electrons from charmed (a) and bottom (b) hadron decays at mid-rapidity. (a) Error bars represent overall uncertainties on simulated \ctoe\ in \dA\ and boxes show combined statistical and systematic uncertainties on the \ctoe\ in \pp~\cite{STAR:NPE:pp200GeV}. 
  (b) Error bars represent statistical uncertainties on measured \eh\ in \dA\ combined with uncertainty on simulated \ctoe\ in \dA, brackets show systematic uncertainties, and boxes show combined statistical and systematic uncertainties on the \btoe\ in \pp~\cite{STAR:NPE:pp200GeV}. 
}
\label{fig:bottom:RdAu}
\end{center}
\end{figure}

Figure \ref{fig:npe:RdAu} shows \rda\ for \eh\ from PHENIX for minimum bias \dA\ collisions. We compare these data to \rda\ for charm quarks, $\rdac$, where we use \eh\ in \pp\ as a baseline. Measurement of \ctoe\ at low \pt\ is unavailable so far, however \ctoe\ dominates \eh\ spectrum for $\pt < 2 \ \gevc$. Our results describes the \eh\ data for $\pt < 2.5 \ \gevc$, which suggest that the enhancement at low \pt\ may be due to initial $k_T$ broadening of charm quarks. Figure \ref{fig:bottom:RdAu} shows nuclear modification factor for electrons from charmed (\rdac, Fig.~\ref{fig:bottom:RdAu}(a)) and bottom (\rdab, Fig.~\ref{fig:bottom:RdAu}(b)) hadron decays. \rdac\ is consistent with unity, which indicates no significant modification due to shadowing and the Cronin effect in the \pt\ range of $4 - 8 \ \gevc$. \rdac\ decreases with \pt\ which may indicate that the simulated \ctoe\ spectrum at high \pt\ is steeper than observed in the data, however such effect is not significant given available precision. \rdab\ shows a moderate enhancement, although the data are comparable with unity within sizable systematic uncertainties and uncertainties due to \pp\ baseline. The enhancement is expected  based on predictions for bottom quark \rda\ in Ref.~\cite{RpA:theory:Vogt} due to the $k_T$ broadening.

\begin{figure}[!htbp]
\begin{center}
\includegraphics[width=.45\textwidth]{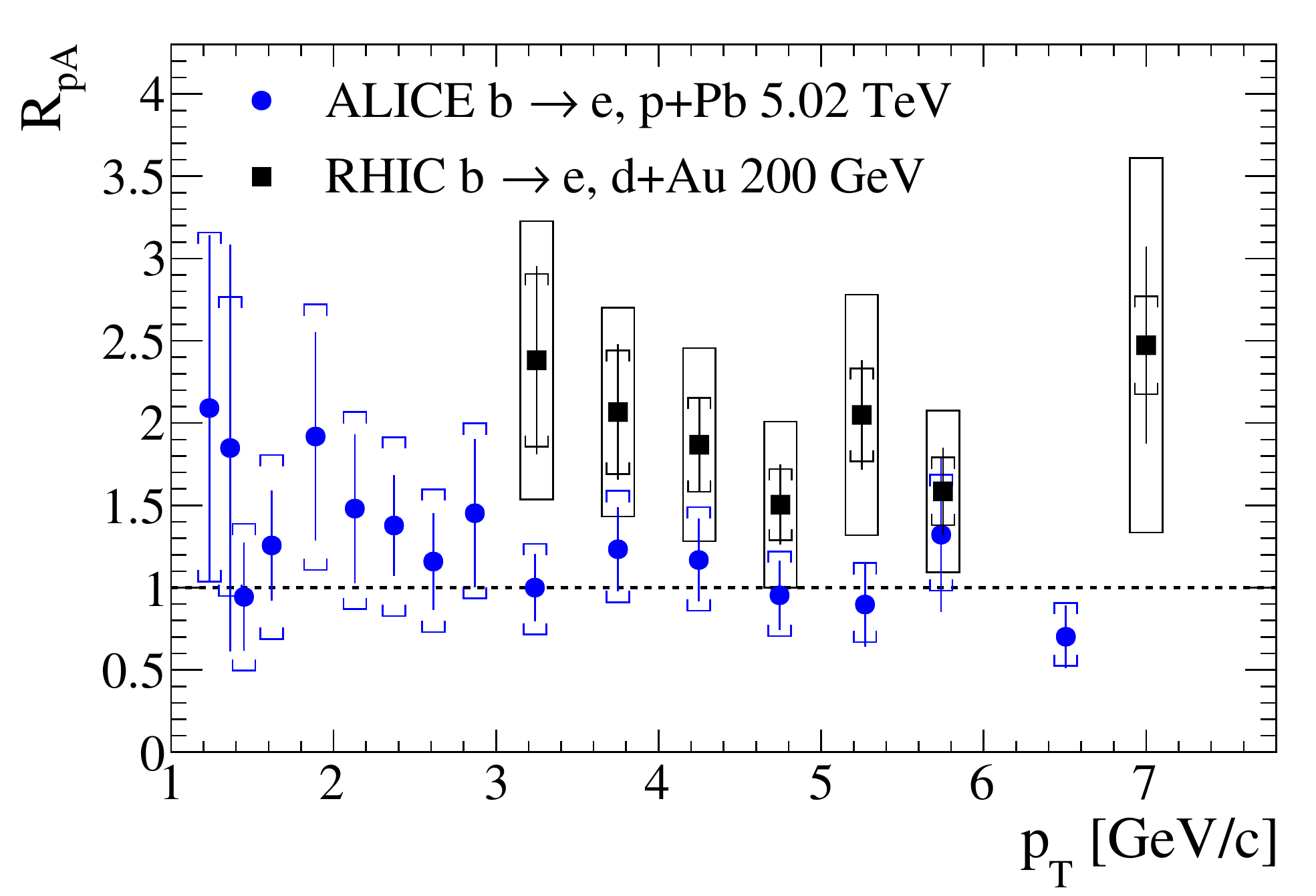}  
\caption{
  (Color online)  Nuclear modification factor for electrons from bottom hadron decays at RHIC ($|y|<0.35$) and LHC ($-1.06 < y < 0.14$). For RHIC results, error bars represent statistical uncertainties on the \eh\ spectra~\cite{Phenix:NPE:dAu} combined with uncertainties on \ctoe\ in \dA, brackets show systematic uncertainties and boxes show combined statistical and systematic uncertainties on the \pp~ baseline~\cite{STAR:NPE:pp200GeV}. In the case of LHC data, error bars represent statistical uncertainties and systematic uncertainties are shown as boxes. 
}
\label{fig:bottom:RHIC:LHC}
\end{center}
\end{figure}

Figure~\ref{fig:bottom:RHIC:LHC} shows $\btoe$ at RHIC compared to LHC data~\cite{ALICE:bottom:QM2014}. Results at RHIC (moderate enhancement for electrons for charmed hadron decays at low-\pt\ and for \btoe\ at higher \pt) are consistent to LHC measurements in \pA\ at $\sqrtsNN = 5.02$~TeV at~\cite{ALICE:bottom:QM2014}); however, LHC data are also consistent with no modification within systematic and statistical uncertainties.

\section{Summary\label{summary}}

We estimated cold nuclear matter effects on beauty production in \dA\ collisions at $\sqrtsNN=200 \ \gev$ at mid-rapidity using as an input recent measurement of the charm \pt\ spectrum in $\sqrts = 200 \ \gev$. Our calculations for \btoe\ show a moderate enhancement for $3 < \pt < 7 \ \gevc$. These results indicate that bottom quark production is not suppressed due to cold nuclear matter effects in \dA\ collisions at RHIC. We also found that shadowing and initial $k_T$ breadboarding for charm quarks due to multiple scattering of incoming partons explain the enhancement of \eh\ yield in \dA\ compared to \pp\ baseline for $\pt < 3\ \gevc$ at this rapidity range. Future measurements of charm and bottom production in \dA\ collisions with new vertex detector are necessary for quantitative understanding of the cold nuclear matter effects for heavy flavor production.

\section{Acknowledgements}

This work was supported in part by the Foundation for Polish Science Grant HOMING PLUS/2013-7/8.

\bibliography{bibligraphy.bib}{}
\bibliographystyle{apsrev}

\clearpage

\end{document}